\documentclass[utf8]{FrontiersinHarvard} 

\usepackage{url,hyperref,microtype,subcaption}
\usepackage[onehalfspacing]{setspace}

\usepackage{xcolor}
\usepackage{ulem}

\def\keyFont{\fontsize{8}{11}\helveticabold }
\def\firstAuthorLast{Blasi {et~al.}} 
\def\Authors{Pasquale Blasi\,$^{1,2*}$ and Elena Amato\,$^{3}$}
\def\Address{$^{1}$Gran Sasso Science Institute (GSSI), Viale Francesco Crispi 7, 67100 L’Aquila, Italy \\
$^{2}$INFN-Laboratori Nazionali del Gran Sasso (LNGS), via G. Acitelli 22, 67100 Assergi (AQ), Italy \\
$^{3}$INAF-Osservatorio Astrofisico di Arcetri, Largo E. Fermi 5, 50125 Firenze, Italy}
\def\corrAuthor{P. Blasi}

\def\corrEmail{pasquale.blasi@gssi.it}

\begin{document}
\onecolumn
\firstpage{1}

\title[Escape]{The problem of escape: a missing bit in the theory of the origin of cosmic rays} 

\author[\firstAuthorLast ]{\Authors} 
\address{} 
\correspondance{} 

\extraAuth{}

\maketitle

\begin{abstract}
The escape of cosmic rays from their sources, as well as from the region surrounding a source, or from the galaxy hosting the sources, is a non-linear process that involves a complex chain of events, often overlooked. On the other hand, these phenomena are responsible for setting the maximum energy in accelerators, shaping the source spectra and determining the conditions for escape from the galaxy hosting the sources, a process that is usually modeled by imposing ad hoc boundary conditions in our equations. Here we discuss some of these phenomena and how they affect the spectra of cosmic rays measured at the Earth.

\tiny
\keyFont{ \section{Keywords:} cosmic rays-diffusion-methods, cosmic rays-ISM, instabilities-ISM, clouds, instabilities,
magnetic fields, plasma physics and astrophysics}
\end{abstract}

\section{Introduction}

Understanding how accelerated particles escape the acceleration region and become cosmic rays (CR) is an essential aspect of the origin of cosmic radiation. For the bulk of Galactic CRs, plausibly accelerated at supernova remnant (SNR) shocks, the role of escape is twofold, in that it likely determines the maximum achievable energy at any time and it impacts the spectrum of particles eventually released in the ISM. As far as the first aspect is concerned, it is generally believed that the upstream escape flux is what creates the current that excites instabilities which in turn lead to magnetic field amplification and larger $p_{\rm max}$. On the other hand, the spectrum of CRs a source releases (hereafter we will refer to this as the source spectrum) is the integration over time of the flux of particles leaving the upstream region at the instantaneous maximum momentum $p_{\rm max}(t)$, plus the particles that from downstream, where they are advected and lose energy adiabatically, match the conditions for leaving the remnant.


The journey of CRs in the immediate vicinity of the source is also non trivial. The CR density in such regions may easily exceed that of the CR sea. Moreover, the CR density gradient is expected to be large and this may lead to the excitation of CR induced plasma instabilities, which confine CRs in the near source regions for times longer than one would estimate based on the average Galactic diffusion coefficient. This phenomenon has been discussed as a source of additional grammage, to be added to the grammage typically accumulated during transport on Galactic scales. On the other hand, near-source grammage might be accumulated also independently of CR induced instabilities if the source is located in a complex environment, such as that of a star cluster, for example, where reduced diffusivity has been inferred from gamma-ray observations.

Finally, we will comment on the regulated escape of CRs from the Galaxy at large, a phenomenon that is typically modeled through the adoption of free escape boundary conditions at some assigned locations (edge of the halo). The transport of Galactic CRs, at least up to energies of order $\sim$TeV, is likely to be dominated by self-confinement, namely by a diffusion coefficient that is due to growth and damping of CR induced instabilities. In particular, we will discuss some recent developments in the investigation of non-linear Landau damping (NLLD) on Galactic CR transport. Some comments on the effects of outflows on galactic scales for the transport of CRs will also be provided.

The article is organized as follows: in \S \ref{sec:snr} we review the role of escaping particles in determining the maximum energy of particles accelerated at a SNR shock and the spectrum these sources release in the Interstellar Medium (ISM); \S \ref{sec:near} is dedicated to the impact of escaping CRs on the near source environment and the potential signatures this process might show in gamma-rays; in \S \ref{sec:galesc} we discuss the subtleties entailed by the current descriptions of galactic transport and the potential consequences of self-regulated CR escape from galaxies on the spectrum of Ultra High Energy Cosmic Rays; \S \ref{sec:sum} is devoted to summary and perspectives.

\section{Escape of accelerated particles from a SNR shock}
\label{sec:snr}
The theory of Diffusive Shock Acceleration (DSA) at newtonian shocks
\cite[]{1977DoSSR.234.1306K,Bell:1978dv} has been very successful in providing an explanation of the power law shape of the spectrum of accelerated particles. The slope of the power law is nearly independent of the microphysics details 
of particle scattering, although these completely determine the maximum energy of the accelerated particles.

When applied to SNR shocks, it was soon recognized that the maximum achievable energy is exceedingly low, unless CRs are able to excite plasma instabilities that lead to enhanced particle scattering \cite[]{Bell:1978dv}. Even accounting for the excitation of resonant streaming instability upstream of the shock, the maximum energy that can be reached at a SNR shock does not exceed 100 TeV \cite[]{Lagage:1983zz}, falling short of the {\it knee}, the maximum proton energy associated to Galactic accelerators,  by at least one order of magnitude. 

The discovery of a non-resonant branch of the streaming instability \cite[]{Bell:2004hhd} was crucial for the field. Given its importance, here we briefly summarize the basic aspects of the process (see \cite{Bell:2004hhd} for a MHD approach to it and \cite{AmatoBlasi2009} for a kinetic approach). In the reference frame of the upstream plasma, the incoming shock front with its quasi-isotropic cloud of accelerated particles (assumed here to be protons) represents a net electric current $J_{CR}=e n_{CR} v_s$, where $v_s$ is the shock speed and $n_{CR}$ is the density of CR with charge $e$. In order to preserve global charge and current neutrality, the plasma reacts with a return current by establishing a small drift between thermal electrons and protons. The instability arises because of this return current, and in its linear phase it grows on scales which are very small compared to the gyroradius of the particles dominating the current, whence the attribute ``non-resonant''. The instability is excited if the following condition is fulfilled:
\begin{equation}
    \frac{J_{CR}(>E)E}{e c} \geq \frac{B_0^2}{4\pi},
    \label{eq:threshold}
\end{equation}
where $B_0$ is the magnitude of the pre-existing magnetic field and $E$ is the energy of the CRs dominating the current. If this condition is met, then the maximum growth occurs at a wavenumber that can be estimated as
\begin{equation}
    k_{\rm max} \approx \frac{4\pi}{c B_0} J_{\rm CR}(>E),
\end{equation}
at a rate
\begin{equation}
    \gamma_{\rm max} = k_{\rm max} v_A,
\end{equation}
where $v_A=B_0/\sqrt{4\pi \rho}$ is the Alfv\'en speed upstream, where the gas density is $\rho$. The non-resonant character of the instability is crucial to determining its importance for particle acceleration. Since $k_{\rm max}\gg r_L^{-1} (E)$ (a condition equivalent to Eq. \ref{eq:threshold}), the magnetic perturbations cannot scatter particles effectively during their phase of exponential growth. The growing magnetic field, $\delta B$, is perpendicular to $B_0$ and exerts a force $\sim (1/c)J_{\rm CR} \times \delta B$ on the background plasma, causing a perpendicular displacement $\Delta x$ that can be easily estimated. The instability plausibly stops growing exponentially when $\Delta x$ becomes comparable with the Larmor radius in the amplified field, a condition equivalent to
\begin{equation}
    \frac{\delta B^2}{4\pi}=\frac{J_{CR}(>E)E}{e c},
    \label{eq:saturation}
\end{equation}
which resembles Eq. \ref{eq:threshold} for the onset of the instability: the instability saturates when the scale $k_{\rm max}^{-1}$ becomes comparable with the Larmor radius of the particles dominating the current, a condition equivalent to reaching equipartition between self-generated magnetic energy density and the energy density carried by the current of escaping particles. 

The role of this instability for particle acceleration in SNRs, where the current is carried by particles escaping the upstream region, was first discussed by \cite{Schure2013}.
Let us assume for simplicity that the spectrum of particles accelerated at the strong SNR shock is $n_{\rm CR}(E)\propto E^{-2}$, for relativistic particles. Defining the acceleration efficiency $\xi_{\rm CR}$ as the fraction of kinetic energy flux that is converted into CR pressure, we can write 
$n_{\rm CR}=\frac{3\xi_{\rm CR}\rho v_s^2}{\Lambda E^2}$, 
where $\Lambda\simeq \ln\left(\frac{E_{\rm max}}{m_p c^2}\right)\sim 10$. Following \cite{Schure2013} and \cite{Cristofari2020}, the maximum energy can be estimated by requiring that the instability has at least $n$ e-folds ($n\sim 5$ is typically assumed) of exponential growth before saturation occurs. The condition for the excitation of the instability can be rewritten as 
\begin{equation}
    M_A>\left( \frac{c \Lambda}{3 \xi_{\rm CR} v_A}\right)^{1/3}\approx 100 \left( \frac{\xi_{\rm CR}}{0.1} \right)^{-1/3} \left( \frac{v_A}{10 \rm km/s} \right)^{-1/3}. 
\end{equation}

\begin{figure}[h!]
\begin{center}
\includegraphics[width=7cm]{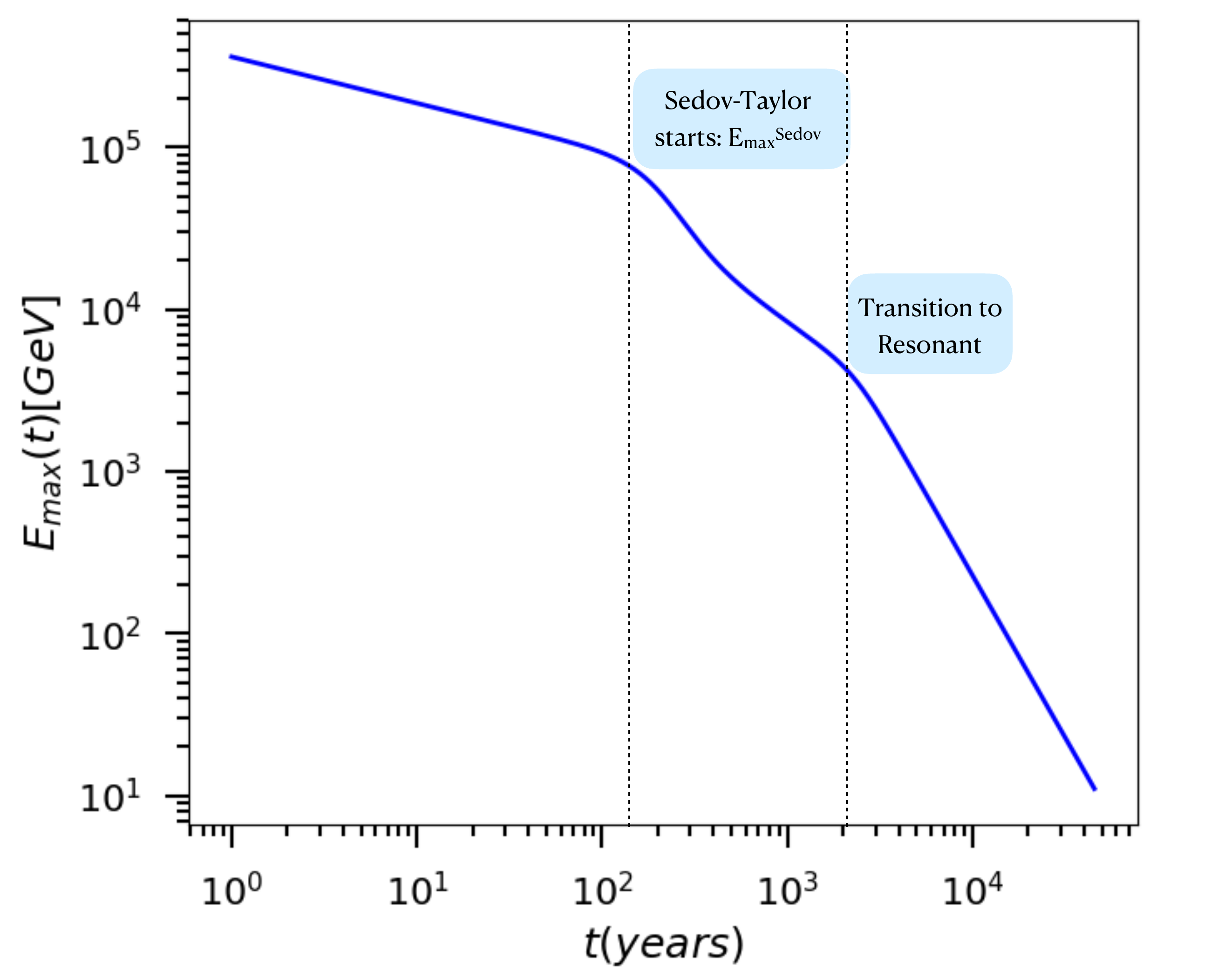}
\includegraphics[width=7cm]{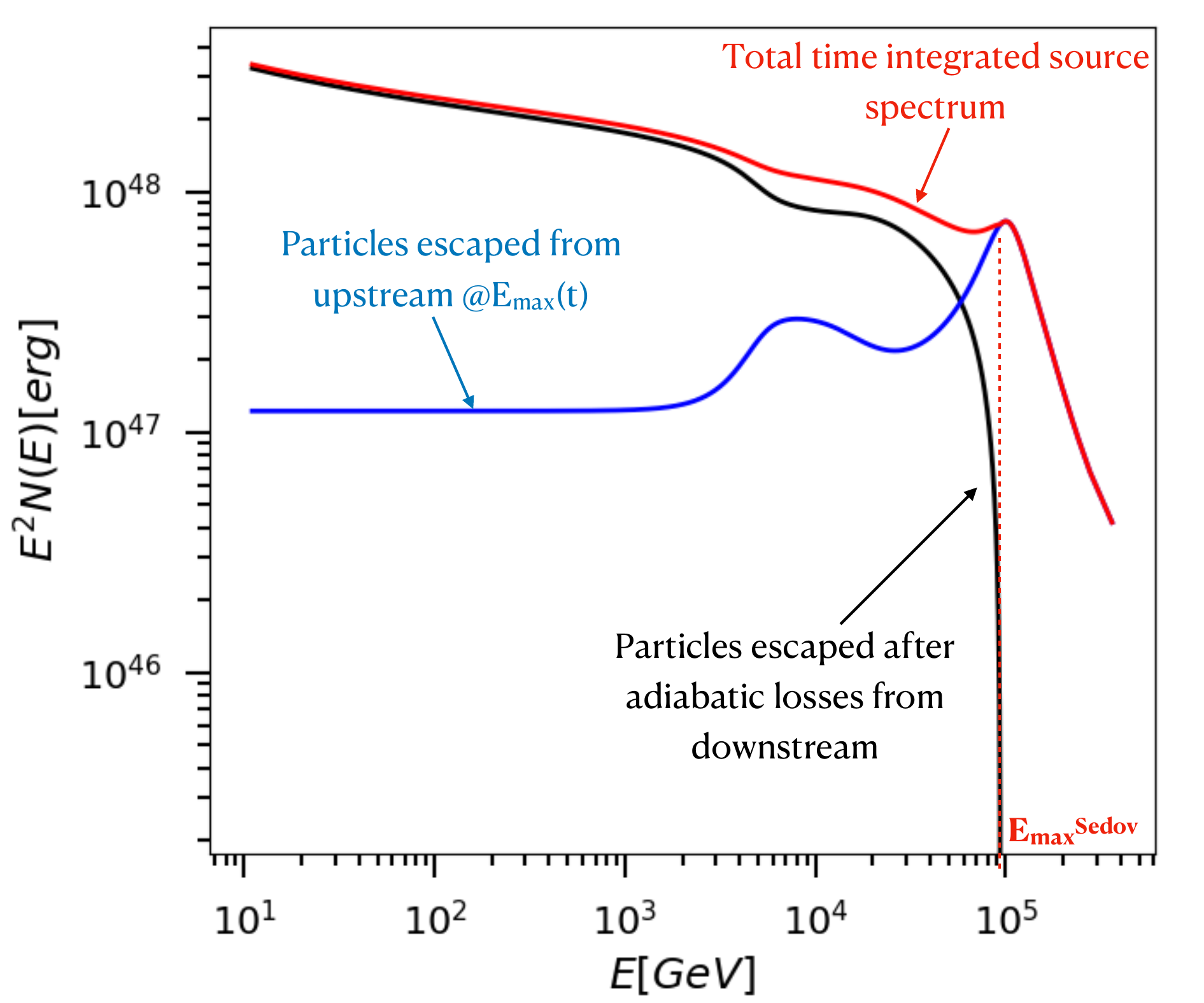}
\end{center}
\caption{{\it Left}: maximum energy as a function of time for parameters typical of a type Ia SNR. {\it Right}: spectrum of particles released into the ISM from upstream (blue line), downstream (black line) and total (red line).}\label{fig:Emax}
\end{figure}

For typical values of $v_A$, this implies that shocks slower than $\sim 1000$ km/s are unable to drive the non resonant instability: for these slower shocks only the resonant instability can lead to magnetic field growth, limited by NLLD (or ion-neutral damping if neutral gas is present). For a type Ia SNR, this leads to $E_{\rm max}(t)$ as shown in the left panel of Fig. \ref{fig:Emax}, where the vertical lines guide the eye to the beginning of the Sedov-Taylor (ST) phase and the time when the non resonant instability stops being excited. 
The non resonant instability grows the fastest for fast shocks in dense environments. 

In principle, this could favour core-collapse SNRs as candidate to achieve the highest energies, since at early times after the explosion these expand in a denser environment (that of the progenitor wind). However, the explosion ejecta are likely more massive in this case, possibly resulting in somewhat lower shock velocity for given explosion energy. For a type Ia SNR the density the shock encounters is assumed to be constant, hence $E_{\rm max}$ is the largest when the SNR is very young (ejecta dominated phase). At these times $E_{\rm max}$ can reach several hundred TeV, yet the amount of matter processed by the shock is small. On the other hand, by the beginning of the ST phase $E_{\rm max}$ has dropped below 100 TeV. When the transition to the resonant SI is reached, $E_{\rm max}\sim 2$ TeV and from then on it quickly further decreases as a result of wave damping. 

The results we have just discussed depend, quantitatively, on the specific assumption one makes for the current of escaping particles, which determines the growth rate of the non resonant streaming instability. We now turn to discuss the spectrum of particles the SNR releases in the ISM:
given the importance of this question, one may think that this is well established but it is not so. There are several recipes that may be adopted and they all give slightly different results, with uncertainties that are larger than the ones associated with current observations in the $\lesssim$TeV energy region. The most common prescription \cite[]{Zira2005} is to assume that escape occurs due to two processes: 1) particles at $E_{\rm max}(t)$ can leave the upstream region at any given time $t$; 2) particles advected downstream lose energy adiabatically and can leave the downstream region when their energy drops below $E_{\rm max}$ at that given time. 

The spectrum of CRs released into the ISM is plotted in the right panel of Fig. \ref{fig:Emax}. A few points need to be made: 1) the time integrated spectrum, namely the source spectrum as we refer to in transport calculations, is not a perfect power law. It shows bumps and dips in the region $\sim 5-30$ TeV, roughly where DAMPE measures a bump in the proton \cite[]{DAMPE2019} and He spectrum \cite[]{Alemanno2021}. The general trend is that of a power law with index slightly steeper than that at the shock itself. 2) The released spectrum (source spectrum hereafter) is dominated by particles escaping from upstream only for very high energies, while the bulk of it is made of particles liberated at later times. 3) The maximum energy that appears in the source spectrum is very close to the $E_{\rm max}$ reached at the beginning of the ST phase. Particles with larger $E_{\rm max}$ end up populating a very steep part of the spectrum (see right panel of Fig. \ref{fig:Emax}). 

In summary, the escaping particles shape the time dependence of $E_{\rm max}(t)$ and by doing so they shape the overall source spectrum after integration over time. As discussed above, there are several prescriptions that one may use to describe particle escape, and they result in somewhat different source spectra, which is worrisome since the source spectrum is a standard input in transport calculations. 

These conclusions also apply to core collapse SNRs, in which case, however, the results are more model dependent since they are affected by the properties of the wind of the progenitor star and the cavity excavated by such wind into the ISM. An instance of this type is Cas A, a $\sim 340$ year old type IIb SNR \cite[]{Krause2008,Rest2011}, located at a distance of 3.4 kpc from the Sun. The current size of the remnant is $\sim 2.8$ pc \cite[]{Vink2022}. The mass of the ejecta associated with the supernova explosion has been estimated as $M_{ej}=2-4~M_\odot$ (\cite{Vink2022} and references therein). The current velocity of the forward shock has been measured to be $\sim 5700-5800$ km/s \cite[]{Vink2022}. These parameters suggest that Cas A is currently at the transition between ejecta dominated and ST phase, a time where it would be expected to accelerate the highest energy particles that would show in its time integrated spectrum. 

LHAASO has recently observed a region of $\sim 2$ degrees around Cas A \cite[]{Cao2024}, which is expected to embed all CRs ever released by Cas A during its history. The gamma ray emission from such region would provide invaluable information about the acceleration history of this SNR. However, only upper limits could be derived to the gamma ray flux in the range $10\leq E_\gamma\leq 1000$ TeV from the region around Cas A, strongly suggesting that the maximum energy of this remnant is currently much below PeV \cite[]{BlasiCasA}.  

Overall, the lack of detection of PeVatron candidates among young SNRs, together with the theoretical difficulties in finding efficient mechanisms for magnetic field amplification casts some serious doubts on the possibility that the bulk of SNRs may be responsible for the CR flux around the knee region. On the theory side, one cannot exclude that some rare more energetic SNRs may be able to provide such high energy CRs, or that other instabilities may enhance the upstream magnetic field to larger values than streaming instabilites can (for instance acoustic instability and turbulent dynamo \cite[]{Drury:1986vd,Drury:2012xb,Beresnyak:2009pi,Capanema26}).

\section{Escape from the near source regions}
\label{sec:near}

Understanding the fate of non thermal particles after they leave the acceleration regions but before they have been fully assimilated into the sea of CRs is crucial for a proper understanding of the origin of CRs, for several reasons: 1) self-confinement in the near source regions, typically in the Galactic disc, may lead to accumulating excess grammage to be added to the one traversed by CRs on their way out of the Galaxy. Such grammage is now measured with high accuracy through the secondary/primary ratios, at least up to a few TV rigidity. In fact, it has been claimed that current B/C measurements allow for at most $\lesssim 0.4~\rm g/cm^2$ excess grammage with respect to the Galactic one \cite[]{Evoli2019}. 2) Suppressed diffusivity around sources may lead to extended regions of production of gamma rays and neutrinos that appear as diffuse but are not accounted for in standard models of diffuse emission \cite[]{Ambrosone2025}. 

The fact that CRs may be self-confined due to the excitation of resonant streaming instability in the strong gradient existing around sources has been discussed by many authors \cite[]{Malkov2013,Nava2016,Dangelo2016,Dangelo2018,Bao2024}. In general these articles agree on the fact that 1) CRs with energy below $\sim$TeV may be confined on scales of few tens of pc around SNRs for times largely exceeding the diffusive times estimated using the Galactic diffusion coefficient; 2) the strength of the effect depends on the abundance of neutral gas, which limits wave-growth due to ion-neutral damping. It is important, however, to keep in mind that the environment around core collapse SNRs is often highly ionized and yet populated with dense molecular clouds. In such an environment the self-confinement can be driven by resonant streaming instability in the ionized gas, while the grammage is accumulated when CRs traverse a cloud. In this situation a sizable grammage can be accumulated in the region around a SNR \cite[]{Bao2024}. During transport in the ionized phase of the ISM, damping is mainly due to NLLD and turbulent damping \cite[]{Farmer2004}. The importance of the latter has however recently been questioned by \cite{Cerri2024}.

The non-linear processes involved in the escape of CRs from their sources can also have a dynamical manifestation: the enhanced CR pressure associated with reduced CR diffusion may lead to the formation of cavities in the circum-source material (see \cite{Schroer2021,Schroer2022} for an investigation of these effects based on hybrid PIC simulations).

CR transport around sources may also be suppressed for reasons that are not directly related to turbulence self-generation: a clear example of this instance is represented by young massive star clusters, regions where a collection of young massive stars, likely blowing powerful winds, is present. If the cluster is compact enough, the individual stellar winds may turn into a collective wind able to evacuate a large region ($\sim 100$ pc) around the cluster. If even a small fraction of the wind kinetic luminosity is converted into a turbulent magnetic field, it is easy to reduce the diffusivity by a few orders of magnitude, with huge implications for gamma ray emission \cite[]{Cesarsky_1983,Morlino_2021,Blasi_2023,Vieu2022,Vieu2023,Vieu2025}.
This holds true even if the cluster is not compact enough for the formation of a collective wind.

As discussed by \cite{Blasi2024} and \cite{Blasi2025}, in these circumstances the grammage accumulated in star clusters is very large and sufficient to cause substantial spallation of nuclei, so that only a fraction of Galactic CRs can actually originate in the environment of 
young massive star clusters. In fact, CR composition \cite[]{Tatischeff21} and direct measurements of gamma-ray emission from star clusters \cite[]{Peron24Nature,PeronM16} suggest that only few \% of CRs originate from these sources.

\begin{figure}[h!]
\begin{center}
\includegraphics[width=7cm]{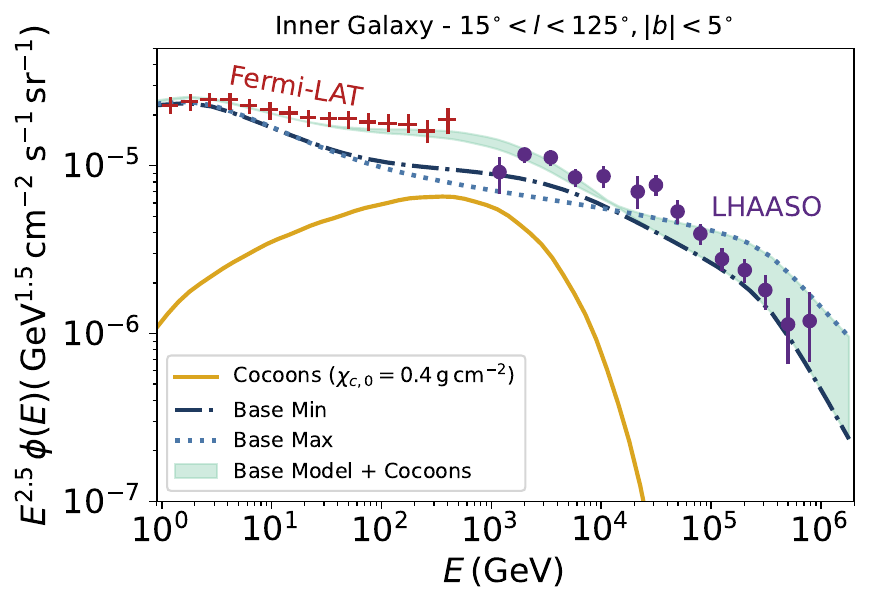}
\includegraphics[width=7cm]{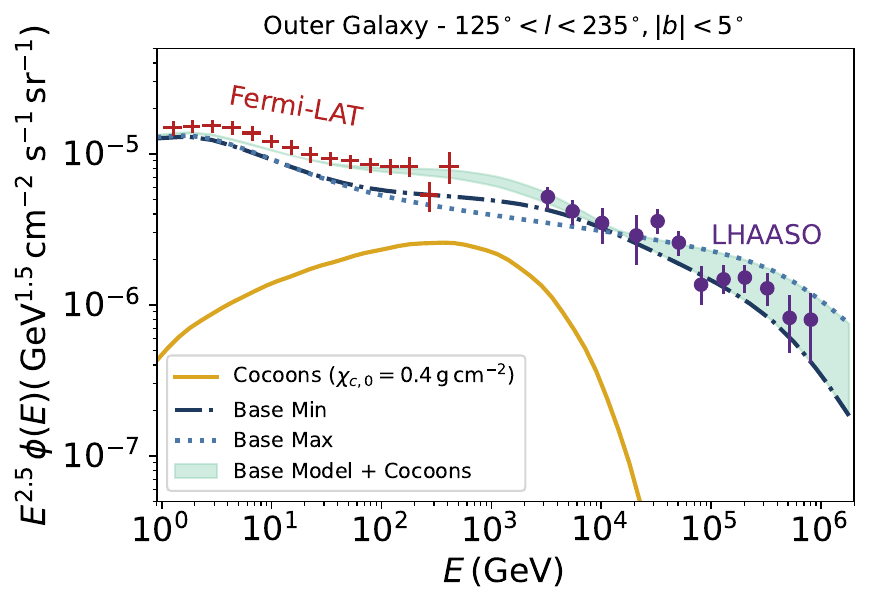}
\end{center}
\caption{{\it Left}: Diffuse gamma ray emission in the inner Galaxy accounting for cocoons where CRs accumulate a grammage $0.4~\rm g/cm^2$ up to a rigidity $R_0$ (here we adopted $R_0=20~\rm TV$). {\it Right}: The same  but for the outer Galaxy. In both panels, the gamma-ray SEDs are compared with Fermi-LAT and LHAASO data reported by~\cite{LHAASO:2024lnz}. Figure adapted from \cite{Ambrosone2025}.}\label{fig:gamma}
\end{figure}

Under milder conditions, it is plausible that these regions may contribute a somewhat smaller amount of grammage and yet their contribution to diffuse gamma-ray and neutrino emission might be sizeable \cite[]{Vecchiotti25}, its entity depending on the assumed transport conditions \cite[]{Menchiari25,Ambrosone2025}.
In Fig. \ref{fig:gamma} we show the Galactic diffuse gamma ray emission in the inner (left panel) and outer (right panel) Galaxy, assuming the existence of regions (the cocoons) where CRs are assumed to accumulate a grammage $0.4~\rm g/cm^2$ (constant in energy with an exponential cutoff at rigidity 20 TV). These fluxes are compared with the ones observed in the same regions by Fermi-LAT and LHAASO \cite[]{LHAASO:2024lnz}. The energy dependence adopted for the cocoons by \cite{Ambrosone2025} may apply to the case of star clusters if the low energy grammage is due to advection. 

It is worth pointing out that recently there have been several claims of suppressed diffusivity around sources: HAWC found evidence for extended TeV halos of gamma ray emission around the Geminga and Monogem pulsars \citep{2017SciHAWK}, and inferred a diffusion coefficient suppressed by more than 2 orders of magnitude compared to the Galactic one. Similar claims have been made for selected SNRs, such as W28 \cite[]{Gabici2010} (see however discussion by \cite{Bao2024} on this instance). The issue of whether these observations suggest CR self-confinement remains open \citep[see e.g. discussion in][]{HaloRev2022Nat,AmatoRecchia2024}.

The increasing instances of sources (such as Cas A) whose surroundings are being observed in gamma rays are going to be keys to a better understanding of the phenomenon of reduced diffusivity.

\section{Escape from galaxies}
\label{sec:galesc}
Self-confinement on Galactic scales and the subsequent escape of CRs from the Galaxy has numerous implications that are only now starting to be actively investigated, although the foundations of the theory of non linear Galactic transport were laid down very early on \cite[]{Holmes1974,Skilling1974}. In that early work it was already clear that, given the density of CRs observed at Earth and for an assumed size of the Galactic halo of few kpc, the gradient should be sufficient to excite resonant streaming instability, at least for moderately high energy CRs. It is rather remarkable that while these ideas were being explored, \cite{1975ApJ...196..107I} first realized that self-generation would result in a CR pressure gradient sufficient, under certain conditions, to overcome gravity and launch a CR driven wind. This idea was later explored in a more realistic setup by \cite{Breit1991,Breit1993}, while \cite{Ptuskin1997} included the transport of CRs in self-generated turbulence in the calculations of the wind properties. It is worth noticing that in these models  the wind launching would occur at some distance from the disc, where the gas density is low enough that the role of ion-neutral damping can be neglected. This introduces a large uncertainty in the properties of the wind and especially on the mass loss rate and the terminal velocity. When an outflow is in fact launched, the CR spectrum that results from the transport equation is typically different from the observed one \cite[]{Recchia2016}. This is due to the role of advection with the wind, that results in a hard spectrum up to $\sim$TeV energies, unless parameters are chosen in an {\it ad hoc} way. 

The calculations of the effect of CRs on outflows (not necessarily winds) in more realistic descriptions of the Galaxy are now carried out using simulations \cite[]{Girichidis2016,Armillotta2024,Hopkins2022,Pfrommer2024}. These provide a powerful tool to bridge the gap between structure formation and the dynamical action of cosmic rays. On the other hand, some of the most problematic aspects of this complex problem lie in the details of microphysics. The most noticeable of the unsettled aspects is the modeling of particle scattering, which governs the coupling between CRs and the ISM and occurs on scales of the order of the particle gyroradius, much smaller than any of the scales accessible to these simulations. A related issue is the description of the wave damping processes, which impact the development of particle self-generated turbulence with consequences on particle scattering and, again, on CR-ISM coupling. The most important of these processes, ion neutral damping and NLLD, both depend on scales not accessible to these simulations, and need a separate investigation with different tools.


In the rest of this section we want to briefly deal with these problems, starting from the connection between self-generation and damping processes. In the standard calculations including self-generation, the CR transport equation is solved with a diffusion coefficient based on quasi-linear theory using the equality between the rate of growth of the streaming instability \cite[]{Kulsrud1969} and the NLLD rate \cite[]{Lee1973}. The non linear transport equation is solved by imposing a free escape boundary condition at a finite distance $H$ from the Galactic disc \cite[]{Blasi2012}, which guarantees that a stationary solution is achieved on time scales longer than the escape time. The latter is normalized to observables such as the secondary/primary ratios and the flux of unstable isotopes. This is exactly what is also done in numerical solutions of the transport equation without self-generation, e.g. by means of codes such as GALPROP \citep{Strong2007,Orlando2018} or DRAGON \citep{Evoli2017, Vittino2018}. 

The assumption of a boundary condition at a fixed surface is highly unsatisfactory conceptually. Most likely, the halo is an emergent structure, that results from processes of self-generation and plasma motion: an attempt to do so, and set the boundary condition of the problem at infinity, was put forward by \cite{Evoli2018}, where the halo was shown to emerge from cascading of pre-existing turbulence and self-generation saturated by NLLD. More recently, in a series of articles, \cite{Dogiel2020} and \cite{Dogiel2022,Dogiel2024} showed how the spectrum of CRs at Earth may develop features quite similar to the ones observed at 300 GV and 20 TV rigidity as a consequence of NLLD and self-generation with boundary conditions at infinity, provided the gas and magnetic fields are properly spatially distributed. The idea is very similar to the one that was first discussed by \cite{Ptuskin1997} in the case of CR driven winds: at large distance from the disc the self-generation becomes prominently more effective than damping, so that the diffusion coefficient drops considerably and particle transport becomes dominated by advection. The distance $z_*(E)$ where the diffusion and advection time become equal plays the role of an effective halo size. In the case with no winds discussed by \cite{Dogiel2020} and \cite{Dogiel2022,Dogiel2024}, the role of the wind velocity is played by the Alfv\'en speed, which depends on location since both the magnetic field and the gas density are stratified. 

As discussed above, a crucial role for the self-generated Galactic CR transport is played by NLLD \cite[]{Lee1973}. The microphysics of the saturation of streaming instability due to NLLD was recently studied by \cite{Schroer2025,Schroer2026}, using hybrid-PIC simulations: this work demonstrated two main facts: 1) the rate of damping is due not only to the waves present at the given wavenumber $k$ but also to all modes with smaller $k$; 2) NLLD results in an inverse cascade that leads to create modes with lower $k$ which can potentially affect scattering of particles with higher energies. Point 1) is especially important for models in which self-generation is only effective up to some energy. In these models, at higher energies, scattering is due to some pre-existing turbulence. \cite{Schroer2026} showed that if such turbulence is present, it contributes excessively to the NLLD of larger $k$ modes so as to drastically reduce the importance of self-generated transport on Galactic scales. On the other hand, models such as those put forward by \cite{Dogiel2020,Dogiel2022} remain viable.

Whether in the context of CR driven winds or in models without outflows, it is clear that the idea that the halo is not a static structure of prescribed size, but rather an emerging one, self-consistently determined by the coupling of CRs and ISM plasma, will play a crucial role in future investigations of the origin of CRs. In all these models a stratification of the large scale magnetic field is either assumed or derived from simulations: this implies that at some distance away from the disc one may expect that the current of CRs crossing a given surface may satisfy the condition for the onset of the non resonant instability (Eq. \ref{eq:threshold}), since eventually the halo has to merge into the standard intergalactic medium. The self-confinement of CRs in regions surrounding our Galaxy (or any galaxy for that matter) was discussed by \cite{AmatoBlasi2019}. This scenario must be an integral part of an approach in which the halo is an emergent structure rather than a fixed well established box. If so, this effect must depend upon the luminosity of the galaxy in the form of CRs and be somewhat more prominent in luminous galaxies. This idea was recently investigated \cite[]{Blasi2015,Cermenati2026} to propose that CRs with energy $\lesssim 10^{18}$ eV may be confined around very luminous galaxies resulting in a flux suppression of Ultra High Energy CRs (UHECRs) at such energies, which is consistent with the very hard injection spectra required by the Auger data \cite[]{abreu2021energy,PAO_AllPart_PhysRevD}. A more quantitative assessment of this agreement requires the calculation of the effect of self-confinement on nuclei, which are also subject to photo-disintegration.

\section{Summary and Conclusions}
\label{sec:sum}
In this article we have provided a short overview of some  outstanding questions and problems related to the escape of energetic particles from their acceleration sites, from the regions around their sources and from galaxies hosting their sources. In all cases discussed above, the current of escaping particles can seed the growth of unstable waves that in turn  ensure efficient scattering. 

In acceleration sites such as Supernova Remnants, this phenomenon is essential to ensure that the maximum energy of accelerated particles reaches levels of astrophysical interest, although it is currently an open question whether the achievable energy is as high as required based upon measured cosmic ray spectra.
On the other hand, the self-confinement of particles near the acceleration site also shapes the spectrum of particles released into the ISM, integrated over the lifetime of the source. This so-called source spectrum is different from the spectrum resulting from DSA and often shows dips and bumps that resemble features recently detected in the spectrum of light CR nuclei. 

We discussed how the escape flux from sources may cause the excitation of unstable waves in the region surrounding them.
This self-confinement around the sources, though unlikely to change the source spectrum as defined above, may result in the accumulation of excess grammage with respect to that usually attributed to CR transport on Galactic scales. The implications of this phenomenon in terms of secondary/primary ratios and in terms of diffuse gamma ray background have been explicitly discussed. 

Finally, we reviewed some recent developments concerning the limitation to the growth of streaming instabilities during Galactic CR transport, as due to NLLD, and how this phenomenon may connect to the physical interpretation of the boundary conditions for the problem of galactic propagation. The escape of CRs from galaxies hosting their physical accelerators may modify the region around these galaxies and lead to the formation of bubbles of suppressed diffusivity and enhanced magnetization. This effect might be at the origin of the UHECR flux suppression detected by Auger at around $10^{18}$ eV.

These topics are at the frontier of investigation nowadays in that we are only now starting to fully realize the far reaching implications of CR self-confinement for a variety of phenomena. More importantly, we now have the computational tools necessary to better study the microphysics of the processes at play and considerably improve their understanding.


\section*{Funding}
This work has been partially funded by the European Union - Next Generation EU under the project IR0000012---CTA+ ``Rafforzamento e creazione di IR nell’ambito del Piano Nazionale di Ripresa e Resilienza (PNRR)'', under PRIN-MUR 2022TJW4EJ ``Unveiling the footprints of the cosmic ray journey through the Galaxy and beyond'' and under MUR National Innovation Ecosystem grant ECS00000041 - VITALITY/ASTRA - CUP D13C21000430001. 




\bibliographystyle{Frontiers-Harvard} 
\bibliography{refs}

\begin{thebibliography}{77}
\providecommand{\natexlab}[1]{#1}
\expandafter\ifx\csname urlstyle\endcsname\relax
  \providecommand{\doi}[1]{doi:\discretionary{}{}{}#1}\else
  \providecommand{\doi}{doi:\discretionary{}{}{}\begingroup
  \urlstyle{rm}\Url}\fi
\providecommand{\selectlanguage}[1]{\relax}
\providecommand{\bibAnnoteFile}[1]{%
  \IfFileExists{#1}{\begin{quotation}\noindent\textsc{Key:} #1\\
  \textsc{Annotation:}\ \input{#1}\end{quotation}}{}}
\providecommand{\bibAnnote}[2]{%
  \begin{quotation}\noindent\textsc{Key:} #1\\
  \textsc{Annotation:}\ #2\end{quotation}}

\bibitem[{Aab et~al.(2020)Aab, Abreu, Aglietta, Albury, Allekotte, Almela
  et~al.}]{PAO_AllPart_PhysRevD}
Aab, A., Abreu, P., Aglietta, M., Albury, J.~M., Allekotte, I., Almela, A.,
  et~al. (2020).
\newblock Measurement of the cosmic-ray energy spectrum above
  $2.5\ifmmode\times\else\texttimes\fi{}{10}^{18}\text{ }\text{ }\mathrm{eV}$
  using the pierre auger observatory.
\newblock \emph{Phys. Rev. D} 102, 062005.
\newblock \doi{10.1103/PhysRevD.102.062005}
\bibAnnoteFile{PAO_AllPart_PhysRevD}

\bibitem[{{Abeysekara} et~al.(2017){Abeysekara}, {Albert}, {Alfaro}, {Alvarez},
  {{\'A}lvarez}, {Arceo} et~al.}]{2017SciHAWK}
{Abeysekara}, A.~U., {Albert}, A., {Alfaro}, R., {Alvarez}, C., {{\'A}lvarez},
  J.~D., {Arceo}, R., et~al. (2017).
\newblock {Extended gamma-ray sources around pulsars constrain the origin of
  the positron flux at Earth}.
\newblock \emph{Science} 358, 911--914.
\newblock \doi{10.1126/science.aan4880}
\bibAnnoteFile{2017SciHAWK}

\bibitem[{Abreu et~al.(2021)Abreu, Aglietta, Albury, Allekotte, Almela,
  Alvarez-Mu{\~n}iz et~al.}]{abreu2021energy}
Abreu, P., Aglietta, M., Albury, J.~M., Allekotte, I., Almela, A.,
  Alvarez-Mu{\~n}iz, J., et~al. (2021).
\newblock The energy spectrum of cosmic rays beyond the turn-down around 10\^{}
  17 10 17 ev as measured with the surface detector of the pierre auger
  observatory.
\newblock \emph{EPJ C} 81, 1--25.
\newblock \doi{https://doi.org/10.1140/epjc/s10052-021-09700-w}
\bibAnnoteFile{abreu2021energy}

\bibitem[{{Alemanno} et~al.(2021){Alemanno}, {An}, {Azzarello}, {Barbato},
  {Bernardini}, {Bi} et~al.}]{Alemanno2021}
{Alemanno}, F., {An}, Q., {Azzarello}, P., {Barbato}, F.~C.~T., {Bernardini},
  P., {Bi}, X.~J., et~al. (2021).
\newblock {Measurement of the Cosmic Ray Helium Energy Spectrum from 70 GeV to
  80 TeV with the DAMPE Space Mission}.
\newblock \emph{Phys. Rev. Lett.} 126, 201102.
\newblock \doi{10.1103/PhysRevLett.126.201102}
\bibAnnoteFile{Alemanno2021}

\bibitem[{{Amato} and {Blasi}(2009)}]{AmatoBlasi2009}
{Amato}, E. and {Blasi}, P. (2009).
\newblock {A kinetic approach to cosmic-ray-induced streaming instability at
  supernova shocks}.
\newblock \emph{MNRAS} 392, 1591--1600.
\newblock \doi{10.1111/j.1365-2966.2008.14200.x}
\bibAnnoteFile{AmatoBlasi2009}

\bibitem[{{Amato} and {Recchia}(2024)}]{AmatoRecchia2024}
{Amato}, E. and {Recchia}, S. (2024).
\newblock {Gamma-ray halos around pulsars: impact on pulsar wind physics and
  galactic cosmic ray transport}.
\newblock \emph{Nuovo Cimento Rivista Serie} 47, 399--452.
\newblock \doi{10.1007/s40766-024-00059-8}
\bibAnnoteFile{AmatoRecchia2024}

\bibitem[{{Ambrosone} et~al.(2025){Ambrosone}, {Evoli}, {Schroer}, and
  {Blasi}}]{Ambrosone2025}
{Ambrosone}, A., {Evoli}, C., {Schroer}, B., and {Blasi}, P. (2025).
\newblock {The origin of very high-energy diffuse {\ensuremath{\gamma}}-ray
  emission: The case for galactic source cocoons}.
\newblock \emph{A\&A} 698, L18.
\newblock \doi{10.1051/0004-6361/202554796}
\bibAnnoteFile{Ambrosone2025}

\bibitem[{{An} et~al.(2019){An}, {Asfandiyarov}, {Azzarello}, {Bernardini},
  {Bi}, {Cai} et~al.}]{DAMPE2019}
{An}, Q., {Asfandiyarov}, R., {Azzarello}, P., {Bernardini}, P., {Bi}, X.~J.,
  {Cai}, M.~S., et~al. (2019).
\newblock {Measurement of the cosmic ray proton spectrum from 40 GeV to 100 TeV
  with the DAMPE satellite}.
\newblock \emph{Science Advances} 5, eaax3793.
\newblock \doi{10.1126/sciadv.aax3793}
\bibAnnoteFile{DAMPE2019}

\bibitem[{{Armillotta} et~al.(2024){Armillotta}, {Ostriker}, {Kim}, and
  {Jiang}}]{Armillotta2024}
{Armillotta}, L., {Ostriker}, E.~C., {Kim}, C.-G., and {Jiang}, Y.-F. (2024).
\newblock {Cosmic-Ray Acceleration of Galactic Outflows in Multiphase Gas}.
\newblock \emph{Ap.J.} 964, 99.
\newblock \doi{10.3847/1538-4357/ad1e5c}
\bibAnnoteFile{Armillotta2024}

\bibitem[{{Bao} et~al.(2024){Bao}, {Blasi}, and {Chen}}]{Bao2024}
{Bao}, Y., {Blasi}, P., and {Chen}, Y. (2024).
\newblock {Regions of Suppressed Diffusion around Supernova Remnants?}
\newblock \emph{Ap.J.} 966, 224.
\newblock \doi{10.3847/1538-4357/ad3939}
\bibAnnoteFile{Bao2024}

\bibitem[{Bell(1978)}]{Bell:1978dv}
Bell, A.~R. (1978).
\newblock {The acceleration of cosmic rays in shock fronts – I}.
\newblock \emph{MNRAS} 182, 147.
\newblock \doi{10.1093/mnras/182.2.147}
\bibAnnoteFile{Bell:1978dv}

\bibitem[{Bell(2004)}]{Bell:2004hhd}
Bell, A.~R. (2004).
\newblock {Turbulent amplification of magnetic field and diffusive shock
  acceleration of cosmic rays}.
\newblock \emph{MNRAS} 353, 550--558.
\newblock \doi{10.1111/j.1365-2966.2004.08097.x}
\bibAnnoteFile{Bell:2004hhd}

\bibitem[{Beresnyak et~al.(2009)Beresnyak, Jones, and
  Lazarian}]{Beresnyak:2009pi}
Beresnyak, A., Jones, T.~W., and Lazarian, A. (2009).
\newblock {Turbulence-induced magnetic fields and the structure of Cosmic Ray
  modified shocks}.
\newblock \emph{Astrophys. J.} 707, 1541--1549.
\newblock \doi{10.1088/0004-637X/707/2/1541}
\bibAnnoteFile{Beresnyak:2009pi}

\bibitem[{{Blasi}(2025{\natexlab{a}})}]{BlasiCasA}
{Blasi}, P. (2025{\natexlab{a}}).
\newblock {Gamma radiation from cosmic rays escaping a young supernova remnant:
  The case of Cas A}.
\newblock \emph{A\&A} 702, A24.
\newblock \doi{10.1051/0004-6361/202555995}
\bibAnnoteFile{BlasiCasA}

\bibitem[{{Blasi}(2025{\natexlab{b}})}]{Blasi2025}
{Blasi}, P. (2025{\natexlab{b}}).
\newblock {Gamma rays from star clusters and implications for the origin of
  Galactic cosmic rays}.
\newblock \emph{A\&A} 694, A244.
\newblock \doi{10.1051/0004-6361/202453017}
\bibAnnoteFile{Blasi2025}

\bibitem[{{Blasi} and {Amato}(2019)}]{AmatoBlasi2019}
{Blasi}, P. and {Amato}, E. (2019).
\newblock {Escape of Cosmic Rays from the Galaxy and Effects on the
  Circumgalactic Medium}.
\newblock \emph{Phys. Rev. Lett.} 122, 051101.
\newblock \doi{10.1103/PhysRevLett.122.051101}
\bibAnnoteFile{AmatoBlasi2019}

\bibitem[{{Blasi} et~al.(2015){Blasi}, {Amato}, and {D'Angelo}}]{Blasi2015}
{Blasi}, P., {Amato}, E., and {D'Angelo}, M. (2015).
\newblock {High-Energy Cosmic Ray Self-Confinement Close to Extra-Galactic
  Sources}.
\newblock \emph{Phys. Rev. Lett.} 115, 121101.
\newblock \doi{10.1103/PhysRevLett.115.121101}
\bibAnnoteFile{Blasi2015}

\bibitem[{{Blasi} et~al.(2012){Blasi}, {Amato}, and {Serpico}}]{Blasi2012}
{Blasi}, P., {Amato}, E., and {Serpico}, P.~D. (2012).
\newblock {Spectral Breaks as a Signature of Cosmic Ray Induced Turbulence in
  the Galaxy}.
\newblock \emph{Phys. Rev. Lett.} 109, 061101.
\newblock \doi{10.1103/PhysRevLett.109.061101}
\bibAnnoteFile{Blasi2012}

\bibitem[{Blasi and Morlino(2023)}]{Blasi_2023}
Blasi, P. and Morlino, G. (2023).
\newblock High-energy cosmic rays and gamma-rays from star clusters: the case
  of cygnus ob2.
\newblock \emph{MNRAS} 523, 4015--4028.
\newblock \doi{10.1093/mnras/stad1662}
\bibAnnoteFile{Blasi_2023}

\bibitem[{{Blasi} and {Morlino}(2024)}]{Blasi2024}
{Blasi}, P. and {Morlino}, G. (2024).
\newblock {Different spectra of cosmic ray H, He, and heavier nuclei escaping
  compact star clusters}.
\newblock \emph{MNRAS} 533, 561--571.
\newblock \doi{10.1093/mnras/stae1782}
\bibAnnoteFile{Blasi2024}

\bibitem[{{Breitschwerdt} et~al.(1991){Breitschwerdt}, {McKenzie}, and
  {Voelk}}]{Breit1991}
{Breitschwerdt}, D., {McKenzie}, J.~F., and {Voelk}, H.~J. (1991).
\newblock {Galactic winds. I. Cosmic ray and wave-driven winds from the
  galaxy.}
\newblock \emph{A\&A} 245, 79
\bibAnnoteFile{Breit1991}

\bibitem[{{Breitschwerdt} et~al.(1993){Breitschwerdt}, {McKenzie}, and
  {Voelk}}]{Breit1993}
{Breitschwerdt}, D., {McKenzie}, J.~F., and {Voelk}, H.~J. (1993).
\newblock {Galactic winds. II. Role of the disk-halo interface in cosmic ray
  driven galactic winds.}
\newblock \emph{A\&A} 269, 54--66
\bibAnnoteFile{Breit1993}

\bibitem[{{Cao} et~al.(2024){Cao}, {Aharonian}, {An}, {Axikegu}, {Bai}, {Bao}
  et~al.}]{Cao2024}
{Cao}, Z., {Aharonian}, F., {An}, Q., {Axikegu}, {Bai}, Y.~X., {Bao}, Y.~W.,
  et~al. (2024).
\newblock {Does or Did the Supernova Remnant Cassiopeia A Operate as a
  PeVatron?}
\newblock \emph{Ap.J. Lett.} 961, L43.
\newblock \doi{10.3847/2041-8213/ad1d62}
\bibAnnoteFile{Cao2024}

\bibitem[{Cao et~al.(2025)}]{LHAASO:2024lnz}
Cao, Z. et~al. (2025).
\newblock {Measurement of Very-High-Energy Diffuse Gamma-Ray Emissions from the
  Galactic Plane with LHAASO-WCDA}.
\newblock \emph{Phys. Rev. Lett.} 134, 081002.
\newblock \doi{10.1103/PhysRevLett.134.081002}
\bibAnnoteFile{LHAASO:2024lnz}

\bibitem[{{Capanema} et~al.(2026){Capanema}, {Blasi}, and
  {Sobacchi}}]{Capanema26}
{Capanema}, A., {Blasi}, P., and {Sobacchi}, E. (2026).
\newblock {Acoustic instability at shockwave precursors}.
\newblock \emph{A\&A} 710, A14.
\newblock \doi{10.1051/0004-6361/202659871}
\bibAnnoteFile{Capanema26}

\bibitem[{{Cermenati} et~al.(2026){Cermenati}, {Aloisio}, {Blasi}, and
  {Evoli}}]{Cermenati2026}
{Cermenati}, A., {Aloisio}, R., {Blasi}, P., and {Evoli}, C. (2026).
\newblock {Excitation of the nonresonant streaming instability around sources
  of ultrahigh-energy cosmic rays}.
\newblock \emph{A\&A} 707, A19.
\newblock \doi{10.1051/0004-6361/202556040}
\bibAnnoteFile{Cermenati2026}

\bibitem[{{Cerri}(2024)}]{Cerri2024}
{Cerri}, S.~S. (2024).
\newblock {Revisiting the role of cosmic-ray driven Alfv{\'e}n waves in
  pre-existing magnetohydrodynamic turbulence. I. Turbulent damping rates and
  feedback on background fluctuations}.
\newblock \emph{A\&A} 688, A182.
\newblock \doi{10.1051/0004-6361/202449492}
\bibAnnoteFile{Cerri2024}

\bibitem[{{Cesarsky} and {Montmerle}(1983)}]{Cesarsky_1983}
{Cesarsky}, C.~J. and {Montmerle}, T. (1983).
\newblock {Gamma-Rays from Active Regions in the Galaxy - the Possible
  Contribution of Stellar Winds}.
\newblock \emph{Sp.Sc.Rev.} 36, 173--193.
\newblock \doi{10.1007/BF00167503}
\bibAnnoteFile{Cesarsky_1983}

\bibitem[{{Chernyshov} et~al.(2022){Chernyshov}, {Dogiel}, {Ivlev}, {Erlykin},
  and {Kiselev}}]{Dogiel2022}
{Chernyshov}, D.~O., {Dogiel}, V.~A., {Ivlev}, A.~V., {Erlykin}, A.~D., and
  {Kiselev}, A.~M. (2022).
\newblock {Formation of the Cosmic-Ray Halo: The Role of Nonlinear Landau
  Damping}.
\newblock \emph{Ap.J.} 937, 107.
\newblock \doi{10.3847/1538-4357/ac8f42}
\bibAnnoteFile{Dogiel2022}

\bibitem[{{Chernyshov} et~al.(2024){Chernyshov}, {Ivlev}, and
  {Dogiel}}]{Dogiel2024}
{Chernyshov}, D.~O., {Ivlev}, A.~V., and {Dogiel}, V.~A. (2024).
\newblock {Secondary cosmic-ray nuclei in the Galactic halo model with
  nonlinear Landau damping}.
\newblock \emph{A\&A} 686, A165.
\newblock \doi{10.1051/0004-6361/202348766}
\bibAnnoteFile{Dogiel2024}

\bibitem[{{Cristofari} et~al.(2020){Cristofari}, {Blasi}, and
  {Amato}}]{Cristofari2020}
{Cristofari}, P., {Blasi}, P., and {Amato}, E. (2020).
\newblock {The low rate of Galactic pevatrons}.
\newblock \emph{Astroparticle Physics} 123, 102492.
\newblock \doi{10.1016/j.astropartphys.2020.102492}
\bibAnnoteFile{Cristofari2020}

\bibitem[{{D'Angelo} et~al.(2016){D'Angelo}, {Blasi}, and
  {Amato}}]{Dangelo2016}
{D'Angelo}, M., {Blasi}, P., and {Amato}, E. (2016).
\newblock {Grammage of cosmic rays around Galactic supernova remnants}.
\newblock \emph{Phys. Rev. D} 94, 083003.
\newblock \doi{10.1103/PhysRevD.94.083003}
\bibAnnoteFile{Dangelo2016}

\bibitem[{{D'Angelo} et~al.(2018){D'Angelo}, {Morlino}, {Amato}, and
  {Blasi}}]{Dangelo2018}
{D'Angelo}, M., {Morlino}, G., {Amato}, E., and {Blasi}, P. (2018).
\newblock {Diffuse gamma-ray emission from self-confined cosmic rays around
  Galactic sources}.
\newblock \emph{MNRAS} 474, 1944--1954.
\newblock \doi{10.1093/mnras/stx2828}
\bibAnnoteFile{Dangelo2018}

\bibitem[{{Dogiel} et~al.(2020){Dogiel}, {Ivlev}, {Chernyshov}, and
  {Ko}}]{Dogiel2020}
{Dogiel}, V.~A., {Ivlev}, A.~V., {Chernyshov}, D.~O., and {Ko}, C.-M. (2020).
\newblock {Formation of the Cosmic-Ray Halo: Galactic Spectrum of Primary
  Cosmic Rays}.
\newblock \emph{Ap.J.} 903, 135.
\newblock \doi{10.3847/1538-4357/abba31}
\bibAnnoteFile{Dogiel2020}

\bibitem[{Drury and Downes(2012)}]{Drury:2012xb}
Drury, L.~O. and Downes, T.~P. (2012).
\newblock {Turbulent magnetic field amplification driven by cosmic-ray pressure
  gradients}.
\newblock \emph{MNRAS} 427, 2308.
\newblock \doi{10.1111/j.1365-2966.2012.22106.x}
\bibAnnoteFile{Drury:2012xb}

\bibitem[{Drury and Falle(1986)}]{Drury:1986vd}
Drury, L.~O. and Falle, S. A. E.~G. (1986).
\newblock {On the Stability of Shocks Modified by Particle Acceleration}.
\newblock \emph{MNRAS} 223, 353.
\newblock \doi{10.1093/mnras/223.2.353}
\bibAnnoteFile{Drury:1986vd}

\bibitem[{{Evoli} et~al.(2019){Evoli}, {Aloisio}, and {Blasi}}]{Evoli2019}
{Evoli}, C., {Aloisio}, R., and {Blasi}, P. (2019).
\newblock {Galactic cosmic rays after the AMS-02 observations}.
\newblock \emph{Phys. Rev. D} 99, 103023.
\newblock \doi{10.1103/PhysRevD.99.103023}
\bibAnnoteFile{Evoli2019}

\bibitem[{{Evoli} et~al.(2018){Evoli}, {Blasi}, {Morlino}, and
  {Aloisio}}]{Evoli2018}
{Evoli}, C., {Blasi}, P., {Morlino}, G., and {Aloisio}, R. (2018).
\newblock {Origin of the Cosmic Ray Galactic Halo Driven by Advected Turbulence
  and Self-Generated Waves}.
\newblock \emph{Phys. Rev. Lett.} 121, 021102.
\newblock \doi{10.1103/PhysRevLett.121.021102}
\bibAnnoteFile{Evoli2018}

\bibitem[{{Evoli} et~al.(2017){Evoli}, {Gaggero}, {Vittino}, {Di Bernardo}, {Di
  Mauro}, {Ligorini} et~al.}]{Evoli2017}
{Evoli}, C., {Gaggero}, D., {Vittino}, A., {Di Bernardo}, G., {Di Mauro}, M.,
  {Ligorini}, A., et~al. (2017).
\newblock {Cosmic-ray propagation with DRAGON2: I. numerical solver and
  astrophysical ingredients}.
\newblock \emph{JCAP} 2017, 015.
\newblock \doi{10.1088/1475-7516/2017/02/015}
\bibAnnoteFile{Evoli2017}

\bibitem[{{Farmer} and {Goldreich}(2004)}]{Farmer2004}
{Farmer}, A.~J. and {Goldreich}, P. (2004).
\newblock {Wave Damping by Magnetohydrodynamic Turbulence and Its Effect on
  Cosmic-Ray Propagation in the Interstellar Medium}.
\newblock \emph{Ap.J.} 604, 671--674.
\newblock \doi{10.1086/382040}
\bibAnnoteFile{Farmer2004}

\bibitem[{{Gabici} et~al.(2010){Gabici}, {Casanova}, {Aharonian}, and
  {Rowell}}]{Gabici2010}
{Gabici}, S., {Casanova}, S., {Aharonian}, F.~A., and {Rowell}, G. (2010).
\newblock {Constraints on the cosmic ray diffusion coefficient in the W28
  region from gamma-ray observations}.
\newblock In \emph{SF2A-2010: Proceedings of the Annual meeting of the French
  Society of Astronomy and Astrophysics}, eds. S.~{Boissier},
  M.~{Heydari-Malayeri}, R.~{Samadi}, and D.~{Valls-Gabaud}. 313.
\newblock \doi{10.48550/arXiv.1009.5291}
\bibAnnoteFile{Gabici2010}

\bibitem[{{Girichidis} et~al.(2016){Girichidis}, {Naab}, {Walch}, {Hanasz},
  {Mac Low}, {Ostriker} et~al.}]{Girichidis2016}
{Girichidis}, P., {Naab}, T., {Walch}, S., {Hanasz}, M., {Mac Low}, M.-M.,
  {Ostriker}, J.~P., et~al. (2016).
\newblock {Launching Cosmic-Ray-driven Outflows from the Magnetized
  Interstellar Medium}.
\newblock \emph{Ap.J. Lett.} 816, L19.
\newblock \doi{10.3847/2041-8205/816/2/L19}
\bibAnnoteFile{Girichidis2016}

\bibitem[{{Girichidis} et~al.(2024){Girichidis}, {Werhahn}, {Pfrommer},
  {Pakmor}, and {Springel}}]{Pfrommer2024}
{Girichidis}, P., {Werhahn}, M., {Pfrommer}, C., {Pakmor}, R., and {Springel},
  V. (2024).
\newblock {Spectrally resolved cosmic rays - III. Dynamical impact and
  properties of the circumgalactic medium}.
\newblock \emph{MNRAS} 527, 10897--10920.
\newblock \doi{10.1093/mnras/stad3628}
\bibAnnoteFile{Pfrommer2024}

\bibitem[{{Holmes}(1974)}]{Holmes1974}
{Holmes}, J.~A. (1974).
\newblock {An energy-dependent confinement model for galactic cosmic rays}.
\newblock \emph{MNRAS} 166, 155--164.
\newblock \doi{10.1093/mnras/166.2.155}
\bibAnnoteFile{Holmes1974}

\bibitem[{{Hopkins} et~al.(2022){Hopkins}, {Butsky}, {Panopoulou}, {Ji},
  {Quataert}, {Faucher-Gigu{\`e}re} et~al.}]{Hopkins2022}
{Hopkins}, P.~F., {Butsky}, I.~S., {Panopoulou}, G.~V., {Ji}, S., {Quataert},
  E., {Faucher-Gigu{\`e}re}, C.-A., et~al. (2022).
\newblock {First predicted cosmic ray spectra, primary-to-secondary ratios, and
  ionization rates from MHD galaxy formation simulations}.
\newblock \emph{MNRAS} 516, 3470--3514.
\newblock \doi{10.1093/mnras/stac1791}
\bibAnnoteFile{Hopkins2022}

\bibitem[{{Ipavich}(1975)}]{1975ApJ...196..107I}
{Ipavich}, F.~M. (1975).
\newblock {Galactic winds driven by cosmic rays.}
\newblock \emph{Ap.J.} 196, 107--120.
\newblock \doi{10.1086/153397}
\bibAnnoteFile{1975ApJ...196..107I}

\bibitem[{{Krause} et~al.(2008){Krause}, {Birkmann}, {Usuda}, {Hattori},
  {Goto}, {Rieke} et~al.}]{Krause2008}
{Krause}, O., {Birkmann}, S.~M., {Usuda}, T., {Hattori}, T., {Goto}, M.,
  {Rieke}, G.~H., et~al. (2008).
\newblock {The Cassiopeia A Supernova Was of Type IIb}.
\newblock \emph{Science} 320, 1195.
\newblock \doi{10.1126/science.1155788}
\bibAnnoteFile{Krause2008}

\bibitem[{{Krymskii}(1977)}]{1977DoSSR.234.1306K}
{Krymskii}, G.~F. (1977).
\newblock {A regular mechanism for the acceleration of charged particles on the
  front of a shock wave}.
\newblock \emph{Akademiia Nauk SSSR Doklady} 234, 1306--1308
\bibAnnoteFile{1977DoSSR.234.1306K}

\bibitem[{{Kulsrud} and {Pearce}(1969)}]{Kulsrud1969}
{Kulsrud}, R. and {Pearce}, W.~P. (1969).
\newblock {The Effect of Wave-Particle Interactions on the Propagation of
  Cosmic Rays}.
\newblock \emph{Ap.J.} 156, 445.
\newblock \doi{10.1086/149981}
\bibAnnoteFile{Kulsrud1969}

\bibitem[{Lagage and Cesarsky(1983)}]{Lagage:1983zz}
Lagage, P.~O. and Cesarsky, C.~J. (1983).
\newblock {The maximum energy of cosmic rays accelerated by supernova shocks}.
\newblock \emph{A\&A} 125, 249--257
\bibAnnoteFile{Lagage:1983zz}

\bibitem[{{Lee} and {V{\"o}lk}(1973)}]{Lee1973}
{Lee}, M.~A. and {V{\"o}lk}, H.~J. (1973).
\newblock {Damping and Non-Linear Wave-Particle Interactions of
  Alfv{\'e}n-Waves in the Solar Wind}.
\newblock \emph{Astrophys. and Sp. Sc.} 24, 31--49.
\newblock \doi{10.1007/BF00648673}
\bibAnnoteFile{Lee1973}

\bibitem[{{Lemoine-Goumard} et~al.(2025){Lemoine-Goumard}, {H{\"a}rer},
  {Mohrmann}, {Bernet}, {Hinton}, {Peron} et~al.}]{Vieu2025}
{Lemoine-Goumard}, M., {H{\"a}rer}, L., {Mohrmann}, L., {Bernet}, R., {Hinton},
  J., {Peron}, G., et~al. (2025).
\newblock {A cosmic-ray loaded nascent outflow driven by a massive star
  cluster}.
\newblock \emph{Nature Communications} 16, 10820.
\newblock \doi{10.1038/s41467-025-65592-4}
\bibAnnoteFile{Vieu2025}

\bibitem[{{L{\'o}pez-Coto} et~al.(2022){L{\'o}pez-Coto}, {de O{\~n}a Wilhelmi},
  {Aharonian}, {Amato}, and {Hinton}}]{HaloRev2022Nat}
{L{\'o}pez-Coto}, R., {de O{\~n}a Wilhelmi}, E., {Aharonian}, F., {Amato}, E.,
  and {Hinton}, J. (2022).
\newblock {Gamma-ray haloes around pulsars as the key to understanding
  cosmic-ray transport in the Galaxy}.
\newblock \emph{Nature Astronomy} 6, 199--206.
\newblock \doi{10.1038/s41550-021-01580-0}
\bibAnnoteFile{HaloRev2022Nat}

\bibitem[{{Malkov} et~al.(2013){Malkov}, {Diamond}, {Sagdeev}, {Aharonian}, and
  {Moskalenko}}]{Malkov2013}
{Malkov}, M.~A., {Diamond}, P.~H., {Sagdeev}, R.~Z., {Aharonian}, F.~A., and
  {Moskalenko}, I.~V. (2013).
\newblock {Analytic Solution for Self-regulated Collective Escape of Cosmic
  Rays from Their Acceleration Sites}.
\newblock \emph{Ap.J.} 768, 73.
\newblock \doi{10.1088/0004-637X/768/1/73}
\bibAnnoteFile{Malkov2013}

\bibitem[{{McIvor} and {Skilling}(1974)}]{Skilling1974}
{McIvor}, I. and {Skilling}, J. (1974).
\newblock {The leakage of cosmic rays from the galaxy}.
\newblock \emph{MNRAS} 167, 49--54.
\newblock \doi{10.1093/mnras/167.1.49}
\bibAnnoteFile{Skilling1974}

\bibitem[{{Menchiari} et~al.(2025){Menchiari}, {Morlino}, {Amato},
  {Bucciantini}, {Peron}, and {Sacco}}]{Menchiari25}
{Menchiari}, S., {Morlino}, G., {Amato}, E., {Bucciantini}, N., {Peron}, G.,
  and {Sacco}, G. (2025).
\newblock {Contribution of young massive stellar clusters to the Galactic
  diffuse {\ensuremath{\gamma}}-ray emission}.
\newblock \emph{A\&A} 695, A175.
\newblock \doi{10.1051/0004-6361/202450621}
\bibAnnoteFile{Menchiari25}

\bibitem[{{Morlino} et~al.(2021){Morlino}, {Blasi}, {Peretti}, and
  {Cristofari}}]{Morlino_2021}
{Morlino}, G., {Blasi}, P., {Peretti}, E., and {Cristofari}, P. (2021).
\newblock {Particle acceleration in winds of star clusters}.
\newblock \emph{MNRAS} 504, 6096--6105.
\newblock \doi{10.1093/mnras/stab690}
\bibAnnoteFile{Morlino_2021}

\bibitem[{{Nava} et~al.(2016){Nava}, {Gabici}, {Marcowith}, {Morlino}, and
  {Ptuskin}}]{Nava2016}
{Nava}, L., {Gabici}, S., {Marcowith}, A., {Morlino}, G., and {Ptuskin}, V.~S.
  (2016).
\newblock {Non-linear diffusion of cosmic rays escaping from supernova remnants
  - I. The effect of neutrals}.
\newblock \emph{MNRAS} 461, 3552--3562.
\newblock \doi{10.1093/mnras/stw1592}
\bibAnnoteFile{Nava2016}

\bibitem[{{Orlando} et~al.(2018){Orlando}, {Johannesson}, {Moskalenko},
  {Porter}, and {Strong}}]{Orlando2018}
{Orlando}, E., {Johannesson}, G., {Moskalenko}, I.~V., {Porter}, T.~A., and
  {Strong}, A. (2018).
\newblock {GALPROP cosmic-ray propagation code: recent results and updates}.
\newblock \emph{Nuclear and Particle Physics Proceedings} 297-299, 129--134.
\newblock \doi{10.1016/j.nuclphysbps.2018.07.020}
\bibAnnoteFile{Orlando2018}

\bibitem[{{Peron} et~al.(2024){Peron}, {Casanova}, {Gabici}, {Baghmanyan}, and
  {Aharonian}}]{Peron24Nature}
{Peron}, G., {Casanova}, S., {Gabici}, S., {Baghmanyan}, V., and {Aharonian},
  F. (2024).
\newblock {The contribution of winds from star clusters to the Galactic
  cosmic-ray population}.
\newblock \emph{Nature Astronomy} 8, 530--537.
\newblock \doi{10.1038/s41550-023-02168-6}
\bibAnnoteFile{Peron24Nature}

\bibitem[{{Peron} et~al.(2025){Peron}, {Menchiari}, {Morlino}, and
  {Amato}}]{PeronM16}
{Peron}, G., {Menchiari}, S., {Morlino}, G., and {Amato}, E. (2025).
\newblock {Hadronic acceleration in the young star cluster NGC 6611 inside the
  M16 region unveiled by Fermi-LAT: Constraints on the acceleration
  efficiency}.
\newblock \emph{A\&A} 703, L8.
\newblock \doi{10.1051/0004-6361/202556564}
\bibAnnoteFile{PeronM16}

\bibitem[{{Ptuskin} et~al.(1997){Ptuskin}, {Voelk}, {Zirakashvili}, and
  {Breitschwerdt}}]{Ptuskin1997}
{Ptuskin}, V.~S., {Voelk}, H.~J., {Zirakashvili}, V.~N., and {Breitschwerdt},
  D. (1997).
\newblock {Transport of relativistic nucleons in a galactic wind driven by
  cosmic rays.}
\newblock \emph{A\&A} 321, 434--443
\bibAnnoteFile{Ptuskin1997}

\bibitem[{{Ptuskin} and {Zirakashvili}(2005)}]{Zira2005}
{Ptuskin}, V.~S. and {Zirakashvili}, V.~N. (2005).
\newblock {On the spectrum of high-energy cosmic rays produced by supernova
  remnants in the presence of strong cosmic-ray streaming instability and wave
  dissipation}.
\newblock \emph{A\&A} 429, 755--765.
\newblock \doi{10.1051/0004-6361:20041517}
\bibAnnoteFile{Zira2005}

\bibitem[{{Recchia} et~al.(2016){Recchia}, {Blasi}, and
  {Morlino}}]{Recchia2016}
{Recchia}, S., {Blasi}, P., and {Morlino}, G. (2016).
\newblock {Cosmic ray driven Galactic winds}.
\newblock \emph{MNRAS} 462, 4227--4239.
\newblock \doi{10.1093/mnras/stw1966}
\bibAnnoteFile{Recchia2016}

\bibitem[{{Rest} et~al.(2011){Rest}, {Foley}, {Sinnott}, {Welch}, {Badenes},
  {Filippenko} et~al.}]{Rest2011}
{Rest}, A., {Foley}, R.~J., {Sinnott}, B., {Welch}, D.~L., {Badenes}, C.,
  {Filippenko}, A.~V., et~al. (2011).
\newblock {Direct Confirmation of the Asymmetry of the Cas A Supernova with
  Light Echoes}.
\newblock \emph{Ap.J.} 732, 3.
\newblock \doi{10.1088/0004-637X/732/1/3}
\bibAnnoteFile{Rest2011}

\bibitem[{{Schroer} et~al.(2025){Schroer}, {Caprioli}, and
  {Blasi}}]{Schroer2025}
{Schroer}, B., {Caprioli}, D., and {Blasi}, P. (2025).
\newblock {Role of Nonlinear Landau Damping for Cosmic-Ray Transport}.
\newblock \emph{Phys. Rev. Lett.} 134, 045201.
\newblock \doi{10.1103/PhysRevLett.134.045201}
\bibAnnoteFile{Schroer2025}

\bibitem[{{Schroer} et~al.(2026){Schroer}, {Caprioli}, and
  {Blasi}}]{Schroer2026}
{Schroer}, B., {Caprioli}, D., and {Blasi}, P. (2026).
\newblock {Investigating nonlinear Landau damping in hybrid simulations}.
\newblock \emph{Phys. Rev. D} 113, 023025.
\newblock \doi{10.1103/cllv-6m95}
\bibAnnoteFile{Schroer2026}

\bibitem[{{Schroer} et~al.(2021){Schroer}, {Pezzi}, {Caprioli}, {Haggerty}, and
  {Blasi}}]{Schroer2021}
{Schroer}, B., {Pezzi}, O., {Caprioli}, D., {Haggerty}, C., and {Blasi}, P.
  (2021).
\newblock {Dynamical Effects of Cosmic Rays on the Medium Surrounding Their
  Sources}.
\newblock \emph{Ap.J.Lett.} 914, L13.
\newblock \doi{10.3847/2041-8213/ac02cd}
\bibAnnoteFile{Schroer2021}

\bibitem[{{Schroer} et~al.(2022){Schroer}, {Pezzi}, {Caprioli}, {Haggerty}, and
  {Blasi}}]{Schroer2022}
{Schroer}, B., {Pezzi}, O., {Caprioli}, D., {Haggerty}, C.~C., and {Blasi}, P.
  (2022).
\newblock {Cosmic-ray generated bubbles around their sources}.
\newblock \emph{MNRAS} 512, 233--244.
\newblock \doi{10.1093/mnras/stac466}
\bibAnnoteFile{Schroer2022}

\bibitem[{{Schure} and {Bell}(2013)}]{Schure2013}
{Schure}, K.~M. and {Bell}, A.~R. (2013).
\newblock {Cosmic ray acceleration in young supernova remnants}.
\newblock \emph{MNRAS} 435, 1174--1185.
\newblock \doi{10.1093/mnras/stt1371}
\bibAnnoteFile{Schure2013}

\bibitem[{{Strong} et~al.(2007){Strong}, {Moskalenko}, and
  {Ptuskin}}]{Strong2007}
{Strong}, A.~W., {Moskalenko}, I.~V., and {Ptuskin}, V.~S. (2007).
\newblock {Cosmic-Ray Propagation and Interactions in the Galaxy}.
\newblock \emph{Annual Review of Nuclear and Particle Science} 57, 285--327.
\newblock \doi{10.1146/annurev.nucl.57.090506.123011}
\bibAnnoteFile{Strong2007}

\bibitem[{{Tatischeff} et~al.(2021){Tatischeff}, {Raymond}, {Duprat}, {Gabici},
  and {Recchia}}]{Tatischeff21}
{Tatischeff}, V., {Raymond}, J.~C., {Duprat}, J., {Gabici}, S., and {Recchia},
  S. (2021).
\newblock {The origin of Galactic cosmic rays as revealed by their
  composition}.
\newblock \emph{MNRAS} 508, 1321--1345.
\newblock \doi{10.1093/mnras/stab2533}
\bibAnnoteFile{Tatischeff21}

\bibitem[{{Vecchiotti} et~al.(2025){Vecchiotti}, {Peron}, {Amato}, {Menchiari},
  {Morlino}, {Pagliaroli} et~al.}]{Vecchiotti25}
{Vecchiotti}, V., {Peron}, G., {Amato}, E., {Menchiari}, S., {Morlino}, G.,
  {Pagliaroli}, G., et~al. (2025).
\newblock {Interpreting the LHAASO Galactic diffuse emission data}.
\newblock \emph{JCAP} 2025, 041.
\newblock \doi{10.1088/1475-7516/2025/09/041}
\bibAnnoteFile{Vecchiotti25}

\bibitem[{{Vieu} et~al.(2022){Vieu}, {Gabici}, {Tatischeff}, and
  {Ravikularaman}}]{Vieu2022}
{Vieu}, T., {Gabici}, S., {Tatischeff}, V., and {Ravikularaman}, S. (2022).
\newblock {Cosmic ray production in superbubbles}.
\newblock \emph{MNRAS} 512, 1275--1293.
\newblock \doi{10.1093/mnras/stac543}
\bibAnnoteFile{Vieu2022}

\bibitem[{{Vieu} and {Reville}(2023)}]{Vieu2023}
{Vieu}, T. and {Reville}, B. (2023).
\newblock {Massive star cluster origin for the galactic cosmic ray population
  at very-high energies}.
\newblock \emph{MNRAS} 519, 136--147.
\newblock \doi{10.1093/mnras/stac3469}
\bibAnnoteFile{Vieu2023}

\bibitem[{{Vink} et~al.(2022){Vink}, {Patnaude}, and {Castro}}]{Vink2022}
{Vink}, J., {Patnaude}, D.~J., and {Castro}, D. (2022).
\newblock {The Forward and Reverse Shock Dynamics of Cassiopeia A}.
\newblock \emph{Ap.J.} 929, 57.
\newblock \doi{10.3847/1538-4357/ac590f}
\bibAnnoteFile{Vink2022}

\bibitem[{{Vittino} et~al.(2018){Vittino}, {Evoli}, {Gaggero}, {Di Bernardo},
  {Di Mauro}, {Ligorini} et~al.}]{Vittino2018}
{Vittino}, A., {Evoli}, C., {Gaggero}, D., {Di Bernardo}, G., {Di Mauro}, M.,
  {Ligorini}, A., et~al. (2018).
\newblock {DRAGON2 : A novel code for Cosmic-Ray transport in the Galaxy}.
\newblock \emph{Nuclear and Particle Physics Proceedings} 297-299, 135--142.
\newblock \doi{10.1016/j.nuclphysbps.2018.07.021}
\bibAnnoteFile{Vittino2018}

\end{thebibliography}




\end{document}